\documentclass[11pt,a4paper]{article}
\usepackage{jcappub}
\usepackage[normalem]{ulem}
\usepackage{xcolor}
\usepackage{bm}
\usepackage{bbm}
\usepackage{amsmath}
\usepackage{etoolbox}
\usepackage{float}
\usepackage{mhchem}
\usepackage{gensymb}
\usepackage{color}
\usepackage{graphicx}
\usepackage{contour}
\usepackage{subcaption}
\usepackage{appendix}
\usepackage{mathtools,leftindex,tensor,mhchem}
\usepackage{tikz}
\usepackage{tikz-feynman}
\usepackage{ulem}
\tikzfeynmanset{compat=1.1.0}
\usetikzlibrary{positioning}
\usetikzlibrary{calc}
\tikzfeynmanset{compat=1.1.0}

\newcommand{\gag}{g_{a\gamma}}

\usepackage{etoolbox}

\makeatletter
\gdef\@fpheader{}
\makeatother

\begin{document}

\makeatletter

\title{NuSTAR as an Axion Helioscope: probing axion-nucleon and axion-electron couplings}

\author[a,b]{Tiziano Zanzarella,}
\author[c,d]{Francisco R. Cand\'on,}
\author[d]{Maurizio Giannotti,}
\author[a,b]{Marco Regis,}
\author[c]{Jaime Ruz,}
\author[b]{Marco Taoso,}
\author[a,e,f]{Elisa Todarello,}
\author[c]{Julia K. Vogel}

\affiliation[a]{Dipartimento di Fisica, Universit\`{a} di Torino, via P. Giuria 1, I--10125 Torino, Italy}
\affiliation[b]{Istituto Nazionale di Fisica Nucleare, Sezione di Torino, via P. Giuria 1, I--10125 Torino, Italy}

\affiliation[c]{Fakult{\"{a}}t f{\"{u}}r Physik. Technische Universit{\"{a}}t  Dortmund. Dortmund, \textit{D-44221}, Germany}

\affiliation[d]{Centro de Astropart{\'i}culas y F{\'i}sica de Altas Energ{\'i}as. University of Zaragoza. Zaragoza, \textit{50009}, Spain}

\affiliation[e]{Leinweber Institute for Theoretical Physics, University of California, Berkeley, CA 94720, U.S.A.}
\affiliation[f]{Theoretical Physics Group, Lawrence Berkeley National Laboratory, Berkeley, CA 94720, U.S.A.}

\emailAdd{tiziano.zanzarella@unito.it}
\emailAdd{marco.regis@unito.it}
\emailAdd{marco.taoso@to.infn.it}
\emailAdd{elisa.todarello@berkeley.edu}

\abstract{
We investigate solar X-ray observations as a probe of axions and axion-like particles. These particles can be produced in the interior of the Sun via the conversion of thermal photons, as well as through processes involving axion-electron and axion-nucleon interactions. The resulting axions can then reconvert into photons in the Sun’s atmospheric magnetic field, generating a signal in the X-ray energy range. In this work, we derive new limits on axions using X-ray observations with the Nuclear Spectroscopic Telescope Array (NuSTAR) during the 2020 solar minimum. In the regime where ALP production is dominated by couplings to electrons or nucleons, we obtain bounds on the product of couplings $g_{ae}\cdot g_{a\gamma}\lesssim 1.1\times10^{-24}\,\rm GeV^{-1}$ and $g_{aN}^{\rm eff}\cdot g_{a\gamma}\lesssim 2.3\times 10^{-19}\,\rm GeV^{-1}$ at 95\% CL, for axion masses $m_a\lesssim10^{-6}\,\rm eV$. These constraints strongly improve current ground-based experimental limits, establishing solar X-ray observations as a powerful and robust method for axion searches.
}

\maketitle


\section{Introduction}
\label{sec:introduction}

Axions are hypothetical pseudoscalar particles that were originally introduced 
to solve the strong CP problem, namely the non-observation of CP violation from QCD interactions~\cite{Pec77a,Pec77b,Wei78,Wil77}.
They originate as pseudo-Goldstone bosons from the spontaneous breaking of the so-called Peccei-Quinn (PQ) symmetry.
The axion decay constant $f_a$, associated with the PQ symmetry breaking, is inversely proportional to the axion mass and also determines the strength of the axion’s interactions with Standard Model (SM) particles, up to model-dependent coefficients, so that large values of $f_a$ correspond to light and very weakly interacting axions.
In broader contexts, other pseudoscalar particles with properties similar to those of axions, often referred to as axion-like particles (ALPs), arise in extensions of the SM, or in String Theory frameworks~\cite{Jaeckel:2010ni}.
ALPs are not necessarily connected to the strong CP problem, and their masses and couplings do not generically follow the relation characteristic of QCD axion models, allowing for a wider parameter space. Therefore, ALPs are usually treated as a generalization of the QCD axion, with masses and couplings taken as effectively independent parameters.
Axions and ALPs are also viable dark matter (DM) candidates and can account for the totality of DM in the Universe in certain regions of parameter space. 
Throughout this paper, we will use the terms “axions” and “ALPs” interchangeably.

\noindent Despite significant experimental efforts, no conclusive evidence for the existence of these particles has been observed,
see e.g.~\cite{Graham:2015ouw,Giannotti:2024xhx} for reviews. 
However, the rich phenomenology of ALPs provides various approaches to constrain their properties. In particular, one of the most promising strategies for axion searches relies on their coupling to photons via the Primakoff effect, i.e.\ the conversion of axions into photons as they propagate through external electromagnetic fields~\cite{Raffelt:1987im,Sikivie:1983ip}.
This mechanism is exploited by helioscope experiments, which search for solar axions produced in the interior of the Sun via their conversion into X-ray photons in strong laboratory magnetic fields.
Among these, the CERN Axion Solar Telescope (CAST) currently sets the most stringent laboratory constraints on the axion-photon coupling over a broad range of masses~\cite{CAST:2024eil}.
Astrophysical X-ray telescopes can also function as axion helioscopes, with axion–photon conversion occurring in the magnetic field of the solar atmosphere rather than in an artificial laboratory field~\cite{Carlson:1995xf}.
A recent realization of this approach was given in Ref.~\cite{Ruz:2024gkl}, which analyzed solar X-ray observations from the Nuclear Spectroscopic Telescope Array (NuSTAR) and derived stringent new limits on the axion–photon coupling. These constraints surpass those from CAST and approach the projected sensitivity of the next-generation helioscope experiment IAXO~\cite{IAXO:2019mpb}.

Ref.~\cite{Ruz:2024gkl} focused exclusively on the axion–photon coupling and considered solar axions produced via photon–axion conversion in the solar core through the Primakoff process.
However, generic ALP models also predict couplings to electrons and nucleons, which can also contribute to the solar axion production.
In this work, we extend the analysis of Ref.~\cite{Ruz:2024gkl} by including these additional production mechanisms.
Using the resulting axion-induced X-ray signal and comparing it with NuSTAR observations, we derive competitive constraints on the axion couplings.

The paper is organized as follows. Section~\ref{sec:signal} describes the production of axions in the Sun and their conversion into X-rays in the magnetic field of the solar atmosphere. In Section~\ref{sec:Dataanalysis}, we present the NuSTAR data and discuss the data analysis.
The bounds on axion-electron and axion-nucleon couplings are derived in Section~\ref{sec:results}. Section~\ref{sec:Conclusion} concludes. In Appendix~\ref{app:Astro Bounds} we compare our limits with other relevant astrophysical bounds.

\section{Axion-photon conversion signal from the Sun}
\label{sec:signal}

\subsection{Axion production in the Sun}
\label{sec:ALPproduction}

\begin{figure}
    \centering
    \includegraphics[width=0.8\linewidth]{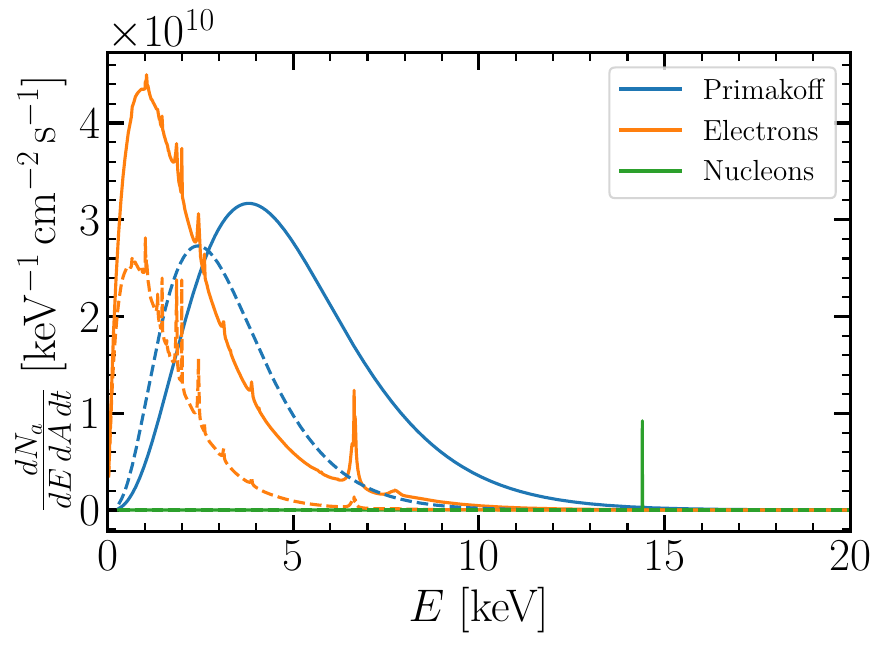}
    \caption{Comparison of axion fluxes originating from a solar disk of radius $0.1\,R_{\odot}$ centered on the Sun (solid lines) and the annulus $0.15\,R_\odot\leq r \leq0.3\,R_\odot$ (dashed lines); respectively our source and background regions. The Primakoff contribution is computed for a coupling $g_{a\gamma}=10^{-10}\,\mathrm{GeV}^{-1}$, the electron-induced component for $g_{ae}=10^{-12}$, and the nucleon-induced contribution for $g_{aN}^{\mathrm{eff}}=10^{-8}$.}
    \label{fig:fluxes}
\end{figure}

\noindent In this section, we discuss ALP production mechanisms in the Sun, focusing on processes relevant at solar thermal energies for each production channel.
In Fig.~\ref{fig:fluxes}, we compare the resulting differential axion fluxes ($\frac{dN_{a}}{dA\,dt\,dE}$), induced by the axion coupling to photons, electrons, and nucleons, as discussed below.
We show the fluxes originated from two regions, one being a circle of radius $0.1\,R_\odot$ centered on the Sun, the other being the annulus $0.15\,R_\odot\leq r \leq0.3\,R_\odot$ (respectively, the source and background regions discussed in Sec.~\ref{sec:nustarobs}).

\subsubsection{Primakoff production}
\noindent The interaction of axions with photons is described by the Lagrangian

\begin{equation}
    \mathcal{L}_{a\gamma}= -\frac{1}{4}g_{a\gamma}aF_{\mu\nu}\tilde{F}^{\mu\nu}\, ,
\end{equation}

\noindent where $a$ is the axion field, $g_{a\gamma}$ parametrizes the effective coupling of axions to photons, and $F_{\mu\nu}$ is the electromagnetic field strength tensor, with $\tilde{F}_{\mu\nu}=\frac{1}{2}\varepsilon_{\mu\nu\rho\sigma}F^{\rho\sigma}$ being its dual. 
The Primakoff effect, that enables the axion-photon conversion in an external electromagnetic field, is a direct consequence of this interaction. 
In the solar interior, thermal photons undergo Primakoff conversion into axions in the presence of the electric fields generated by electrons and ions in the plasma.
The resulting axion flux can be computed with good precision, see~\cite{Hoof:2021mld} for a study of the associated uncertainties. 
It peaks at energies $\sim4\,\rm keV$ and is mainly concentrated in the solar core.
In our analysis, we use the axion flux from the same model employed in~\cite{Hoof:2021mld}

\subsubsection{Production from axion-electron coupling}

\begin{figure}[htbp]
\centering
\captionsetup[subfigure]{labelformat=empty}

\begin{subfigure}{0.48\textwidth}
\centering
\begin{tikzpicture}[scale=1.4, transform shape]
\begin{feynman}
\vertex (a) at (0,0) {$e^-$};
\vertex (b) at ($(a)+(1cm,0)$);
\vertex (c) at ($(a)+(1.5cm,0)$);
\vertex (d) at ($(a)+(0,-1)$) {$e^-$};
\vertex (e) at ($(d)+(1cm,0)$);
\vertex (f) at ($(d)+(2cm,0)$) {$e^-$};
\vertex (h) at ($(c)+(0.5cm,0.5cm)$) {$a$};
\vertex (k) at ($(c)+(0.5cm,-0.5cm)$) {$e^-$};
\diagram*{
(a) -- [fermion] (b) -- [fermion] (c),
(c) -- [fermion] (k),
(c) -- [scalar] (h),
(d) -- [fermion] (e) -- [fermion] (f),
(b) -- [photon, edge label=$\gamma$] (e)
};
\end{feynman}
\end{tikzpicture}
\caption{Electron bremsstrahlung}
\end{subfigure}
\hspace{-2.5cm}
\begin{subfigure}{0.48\textwidth}
\centering
\begin{tikzpicture}[scale=1.4, transform shape]
\begin{feynman}
\vertex (a) at (0,0) {$e^-$};
\vertex (b) at ($(a)+(1cm,0)$);
\vertex (c) at ($(a)+(1.5cm,0)$);
\vertex (d) at ($(a)+(0,-1)$) {$I^+$};
\vertex (e) at ($(d)+(1cm,0)$);
\vertex (f) at ($(d)+(2cm,0)$) {$I^+$};
\vertex (h) at ($(c)+(0.5cm,0.5cm)$) {$a$};
\vertex (k) at ($(c)+(0.5cm,-0.5cm)$) {$e^-$};
\diagram*{
(a) -- [fermion] (b) -- [fermion] (c),
(c) -- [fermion] (k),
(c) -- [scalar] (h),
(d) -- [fermion] (e) -- [fermion] (f),
(b) -- [photon, edge label=$\gamma$] (e)
};
\end{feynman}
\end{tikzpicture}
\caption{Ion bremsstrahlung}
\end{subfigure}

\vspace{0.4cm}

\begin{subfigure}{0.32\textwidth}
\centering
\begin{tikzpicture}[scale=1.4, transform shape]
\begin{feynman}
\vertex (a) at (0,0) {$e^-$};
\vertex (b) at ($(a)+(1cm,0)$);
\vertex (c) at ($(a)+(2cm,0)$) {$a$};
\vertex (d) at ($(a)+(0,-1cm)$) {$\gamma$};
\vertex (e) at ($(d)+(1cm,0)$);
\vertex (f) at ($(e)+(1cm,0)$) {$e^-$};
\diagram* {
(a) -- [fermion] (b) -- [scalar] (c),
(d) -- [photon] (e) -- [fermion] (f),
(b) -- [fermion, edge label=$e^-$] (e)
};
\end{feynman}
\end{tikzpicture}
\caption{Axion compton effect}
\end{subfigure}
\hfill
\begin{subfigure}{0.32\textwidth}
\centering
\begin{tikzpicture}[scale=1.4, transform shape]
\begin{feynman}
\vertex (a) at (0,0) {$I^+$};
\vertex (b) at ($(a)+(1cm,0)$);
\vertex (c) at ($(a)+(2cm,0)$) {$I$};
\vertex (e) at ($(a)+(0,1cm)$) {$e^-$};
\vertex (f) at ($(c)+(0,1cm)$) {$a$};
\diagram*{
(a) -- [fermion] (b) -- [fermion] (c),
(e) -- [fermion] (b) -- [scalar] (f),
};
\end{feynman}
\end{tikzpicture}
\caption{Axio-recombination}
\end{subfigure}
\hfill
\begin{subfigure}{0.32\textwidth}
\centering
\begin{tikzpicture}[scale=1.4, transform shape]
\begin{feynman}
\vertex (a) at (0,0) {$I^*$};
\vertex (b) at ($(a)+(1cm,0)$);
\vertex (c) at ($(b)+(1cm,0.5cm)$) {$I$};
\vertex (d) at ($(b)+(1cm,-0.5cm)$) {$a$};
\diagram*{
(a) -- [fermion] (b),
(b) -- [fermion] (c),
(b) -- [scalar] (d)
};
\end{feynman}
\end{tikzpicture}
\caption{Axio-deexcitation}
\end{subfigure}

\caption{Diagrams of axion-electron interactions (BCA processes) that arise from the Lagrangian in Eq.~\ref{eq:axion electron lagrangian}. $I$ represents an arbitrary atom, with $I^*$ and $I^+$ indicating its excited and singly-ionized states, respectively.
}
\label{fig:BCA processes}

\end{figure}
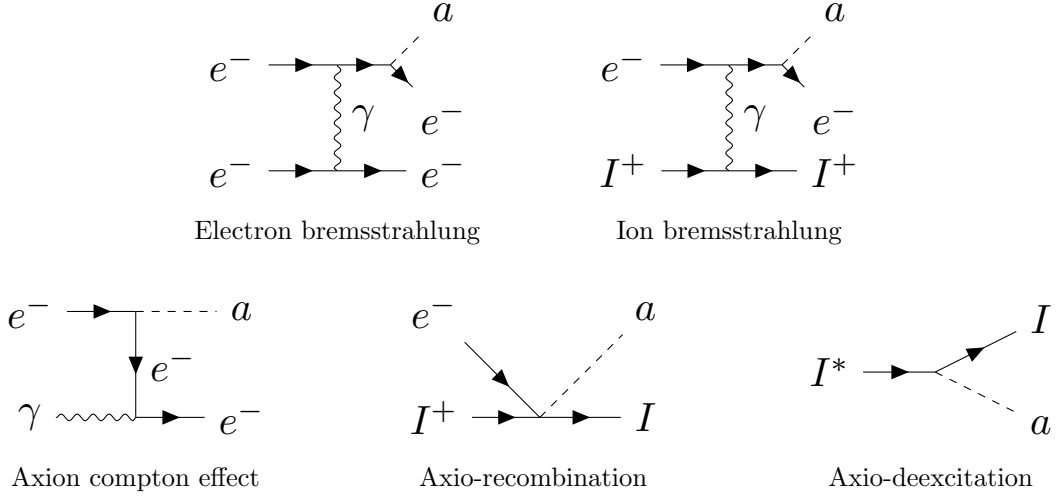

\noindent The interaction of ALPs with fermions can be described by the following Lagrangian

\begin{equation}
\label{eq:axion electron lagrangian}
    \mathcal{L}_{af}= \sum_f \frac{g_{af}}{2m_f}\left(\partial_\mu a\right)\bar{f}\gamma^\mu\gamma^5f\, ,
\end{equation}

\noindent where $g_{af}$ is the coupling constant of axions to fermions and $m_f$ is the fermion mass. 
In the case of electrons, $f\equiv e$, 
the relevant channels for the production of ALPs in stellar systems
are axion bremsstrahlung, axion Compton effect, axio-deexcitation and axio-recombination (BCA processes), represented in Figure~\ref{fig:BCA processes}. 

\noindent In our analysis we use the solar axion flux from~\cite{Redondo_2013,Hoof:2021mld}, which includes all the BCA processes described above, see~\cite{Hoof:2021mld} for an assessment of the associated uncertainties.

\subsubsection{Production from axion-nucleon coupling}
\label{sec:axion-nucleon}

From the Lagrangian in Eq.~(\ref{eq:axion electron lagrangian}), one may also expect nucleonic counterparts of the BCA processes, involving protons and neutrons and mediated by the couplings $g_{ap}$ and $g_{an}$. Their relevance, however, is highly environment dependent. In dense and hot systems such as supernovae, proto-neutron stars, and neutron stars, nucleon--nucleon bremsstrahlung is typically one of the dominant production channels. In the solar interior this channel is strongly suppressed because the nucleons are cold on the nuclear scale: $T_{\rm core}\sim 1.3\,{\rm keV}$ is many orders of magnitude below the characteristic momentum scale of the nuclear interaction, often represented in this context by one-pion exchange. Consequently, the phase space and matrix elements relevant for efficient nucleon--nucleon axion bremsstrahlung are highly suppressed under solar conditions. Solar axion production through nucleon couplings is therefore mainly tied to nuclear processes, including both thermally populated nuclear transitions and non-thermal reactions in the solar  fusion chain. For the range of energies considered here, the most relevant standard contribution is the M1 magnetic transition of $\prescript{57}{}{\rm Fe}$, which can emit an axion at the excitation energy $E^*=14.4\,{\rm keV}$.

\noindent The resulting axion flux at Earth is from this channel is~\cite{Carenza:2024ehj}:

\begin{equation}
\label{eq:Fe flux normalization}
    \Phi_{\prescript{57}{}{\rm Fe}}=5.06\cdot{10}^{23}\, \left(g_{aN}^{\rm eff}\right)^2\, {\rm s}^{-1}\,{\rm cm}^{-2}\, 
\end{equation}

\noindent where, $g_{aN}^{\rm eff}$ is a combination between the axion-proton coupling $g_{ap}$ and the axion-neutron coupling $g_{an}$, expressed as

\begin{equation}
    g_{aN}^{\rm eff}=0.15\, g_{ap}+1.15\,g_{an}\, .
\end{equation}

\noindent The monochromatic axion flux at $14.4\, \rm keV$ undergoes a thermal broadening as axions are emitted from the Sun. Hence, we model the axion signal as a narrow gaussian centered at 14.4 keV, with a spread given by

\begin{equation}
    \sigma_{\rm th}=E^*\sqrt{\frac{T_{\rm core}}{m_{\rm Fe}}}\simeq2\, \rm eV\,,
\end{equation}

\noindent where $E^*=14.4\,\rm{keV}$ is the energy of the spectral line and $m_{\rm Fe}=5.7\cdot10^{7}\,\rm keV$ is the mass of the $\prescript{57}{}{\rm Fe}$ nucleus. The resulting differential axion flux emitted from the Sun core is

\begin{equation}
    \frac{dN_{a,\prescript{57}{}{\rm Fe} }}{dA\, dt\, dE}\left(E\right)=\Phi_{\prescript{57}{}{\rm Fe}}\cdot\frac{1}{\sqrt{2\pi}\sigma_{\rm th}}e^{-\frac{\left(E-E^*\right)^2}{2\sigma_{\rm th}^2}}\,.
\end{equation}

\noindent The $^{57}{\rm Fe}$ axion emissivity is strongly concentrated in the solar core because the thermal population of the $14.4\,{\rm keV}$ excited state is exponentially sensitive to the local temperature. As shown in Ref.~\cite{DiLuzio:2021qct} (specifically, in Fig. 1), the resulting flux is mostly produced within $r\lesssim0.1\,R_\odot$. We therefore can safely neglect the contribution from the region beyond $0.1\,R_{\odot}$.

\subsection{Axion conversion in the solar atmospheric magnetic field}
\label{sec:ALPconversion}

Once they have exited the photosphere, axions produced in the solar core can convert into X-ray photons, which in turn propagate out of the solar atmosphere and reach Earth. The probability of conversion of an axion into a photon as a function of the altitude above the solar surface  $h$ is~\cite{Raffelt:1987im}
\begin{equation}
\label{eq:prob_a_gamma}
P_{a\rightarrow\gamma}(h) = \frac{1}{4}\gag^2\Big|\int_0^h dh' B_\perp(h')\ e^{i\int_0^{h'}dh''q(h'')}\ e^{-\frac{1}{2}\int_{h'}^h dh''\Gamma(h'')}
\Big|^2 ,\enspace 
\end{equation}
where $B_\perp$ is the component of the solar magnetic field perpendicular to the direction of propagation of the axion and photon, $q(h)$ is the momentum mismatch between axion and photon, and $\Gamma(h)$ is the photon absorption coefficient. For highly relativistic axions, the momentum mismatch can be expressed as
$q(h) = (\omega_{p}^2(h) - m_a^2)/2E$, where $E$ is the photon energy, $m_a$ the axion mass and $\omega_{p}$ the plasma frequency, which acts as an effective photon mass. 

To model the plasma and magnetic field of the solar atmosphere, we use an approach completely analogous to that of Ref.~\cite{Ruz:2024gkl}. In our model, we include free electrons and elements up to atomic number $Z=30$. The spatial profiles are taken from \cite{2008GeofI..47..197D}, while the abundances of elements with $Z\geq 2$ are taken from the CHIANTI database. All elements are assigned the same spatial profile as the hydrogen. For the purpose of calculating the momentum mismatch, only free electrons, hydrogen, and helium are relevant, given the small abundance of heavier elements. For the absorption coefficient, heavier elements might be relevant, despite their scarcity, due to their larger photon cross-section. We take the photon cross-sections from the NIST XCOM database~\cite{xcom}, including coherent and incoherent scattering, and photoelectric absorption.

The magnetic field of the photosphere is taken from the magneto-convection simulations with small-scale dynamo activity of Ref.~\cite{Rempel_2014}. We use this model up to an altitude of  400~km above the solar surface. The coronal magnetic field is instead taken from the model developed by Predictive Science Inc. (PSI) MHD simulation corresponding to the solar eclipse of July 2, 2019~\cite{ps2017}. This model covers altitudes between 0.1 and 29 solar radii. We interpolate between the two regions using a power law. We further validate this model by comparing it with Potential-Field-Source-Surface (PFSS) model for the specific day of our observation~\cite{pffs6}, finding excellent agreement in the overlap region (see Fig. 2 of~\cite{Ruz:2024gkl}).

The result of integrating Eq.~\eqref{eq:prob_a_gamma} with the model described above is shown in Fig. D of ~\cite{Ruz:2024gkl}. The conversion probability at $29R_\odot$ saturates for axion masses of order $10^{-7}$~eV and lower, while for higher masses, growth beyond the resonance location ($m_a=\omega_p$) is suppressed by decoherence.

\section{X-ray data and analysis}
\label{sec:Dataanalysis}

\subsection{NuSTAR observations}
\label{sec:nustarobs}

For the analysis presented in this work, we focus on the NuSTAR quiet-Sun campaign of 21-Feb-2020, which provides the most suitable dataset currently available for solar axion searches. This campaign consists of a long dwell at the solar disk center during a low-activity period near solar minimum, with nine spacecraft orbits yielding exposures of 23.8~ks and 24.9~ks for telescope modules $A$ and $B$, respectively. Other observations performed during the same 2018-2020 solar minimum exist, but they were obtained as part of mosaics covering the full solar disk and add only $\sim100$~s of useful exposure. With the next solar minimum expected around 2030, the present dataset is likely to remain the best X-ray satellite dataset for solar axion studies in the near future.

\noindent The 21-Feb-2020 observations were collected between 04:51 and 22:41 GMT, with each orbit providing roughly one hour of exposure. The $10'$ square field of view covers about $18\%$ of the solar disk. 
As the Sun drifted slowly across the field of view during the day, each orbit was divided into 10--15~min segments and the solar center coordinates were updated accordingly. Data products were then extracted separately for each segment. Modules A and B were analyzed independently, because of different time exposure, effective area, and detector response.
Data reduction was performed with the \texttt{NuSTARDAS} package (v2.1.4) within the \texttt{HEASoft} framework (v6.34), using calibration files from \texttt{CALDB} v20240325~\cite{Perri2020}. 
For more details, see Ref.~\cite{Ruz:2024gkl}.

Source and background regions were defined independently for each time segment. The source region is a circle centered on the solar disk with radius $0.1\,R_\odot$ (1.6$'$), while the background region is an annulus spanning $0.15$--$0.30\,R_\odot$. The dominant background component is "stray light", namely diffuse cosmic X-rays reaching the detector without passing through the optics and producing a characteristic gradient across the focal plane due to shadowing by the spacecraft structure~\cite{Wik_2014}. We modeled this contribution with the \texttt{nuskybgd} package and used \texttt{XSPEC} v12.13.1~\cite{arnaud1999xspec} for the final spectral analysis. 

A key difference with respect to the analysis presented in Ref.~\cite{Ruz:2024gkl} is that here we extend the explored energy range up to $15\,\rm keV$. This wider interval allows us to probe the harder part of the expected axion-induced spectra, which is particularly relevant for axion production channels involving nucleon couplings. Although the instrumental background increases at higher energies and the contribution from stray light becomes more significant, the improved background modeling provided by \texttt{nuskybgd} enables a reliable characterization of the detector response over the full $4$–$15\,\rm keV$ range. This extension therefore enhances the sensitivity to axion scenarios predicting non-negligible high-energy photon fluxes.

\begin{figure}[H]
    \centering
    \includegraphics[width=0.75\linewidth]{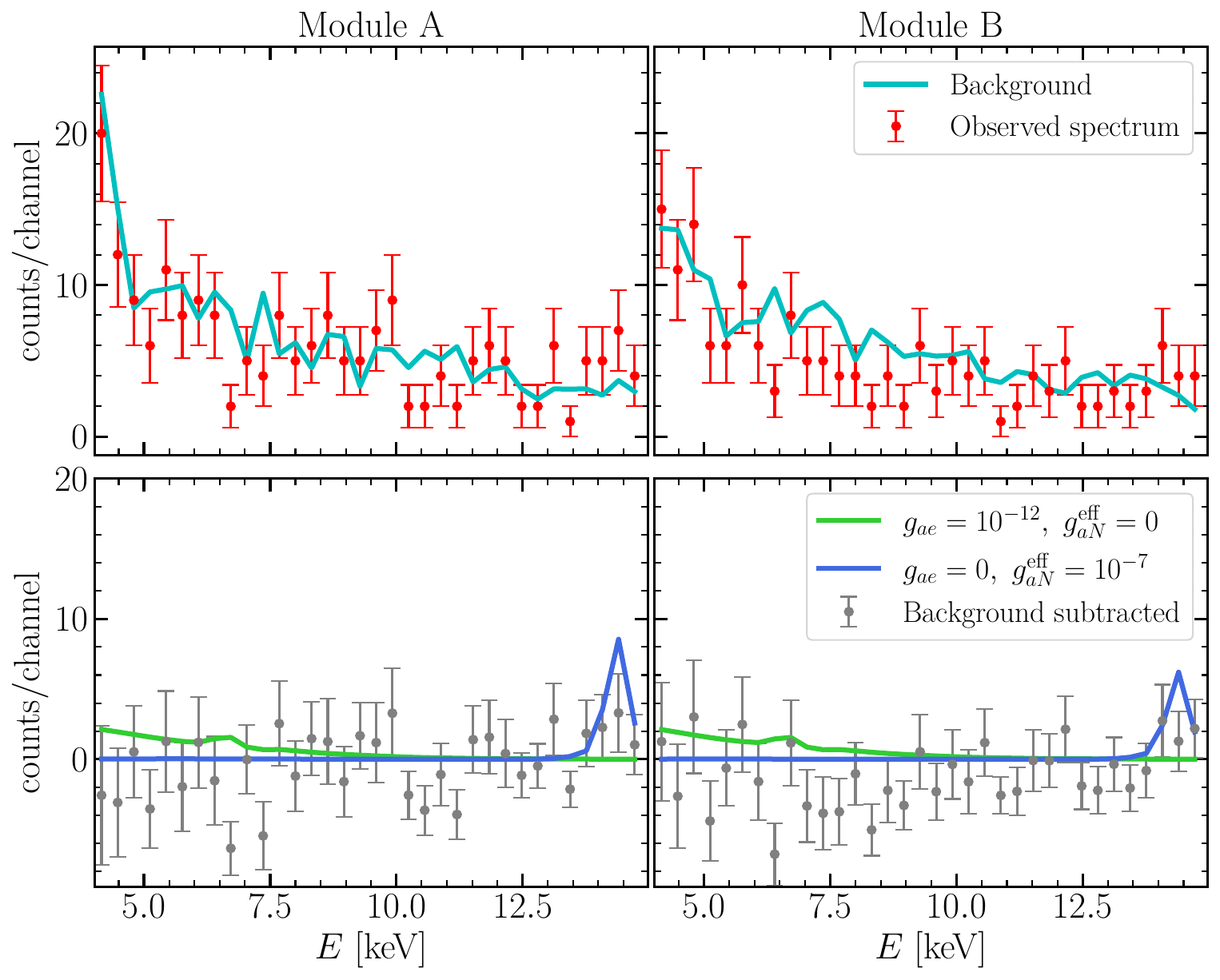}
    \caption{(\textbf{Top row})~NuSTAR counts in the source region (red) and in the background region (cyan), the latter rescaled to the source area, for module A (\textbf{left}) and module B (\textbf{right}). (\textbf{Bottom row}) Background subtracted counts in the source region (gray dots) and 
    the photon signal from axion-electron (green line) and axion-nucleon (blue line) couplings, and for an axion-photon coupling corresponding to the $95\%$ CL limit. 
    These lines are relative to an axion mass of $m_{a}=10^{-8}\, \rm eV$.
    }
    \label{fig:plot counts}
\end{figure}

\noindent In the top panels of Fig.~\ref{fig:plot counts}, we show the photon counts in the source region
over the $4$–$15\,\rm keV$ energy range
for the two modules (red dots),
along with the background counts in the same region (cyan line), obtained by rescaling the counts in the background region to the source area.
The two bottom panels show the background-subtracted signal (gray dots), and the expected photon signal
arising from the conversion of axions produced via axion-electron coupling (green line), and axion-nucleon coupling (blue line). These lines are shown for 
$g_{ae}=10^{-12}$ and $g_{aN}^{\rm eff}=10^{-7}$,
while $g_{a\gamma}$ is fixed to the $95\%$ CL bound (cfr Figs.~\ref{fig:Bounds_experimental_gae},~\ref{fig:Bounds_experimental_gaN}).
For clarity, this plot uses a bin width of $\Delta E=320\,\rm eV$ instead of the $\Delta E=80\,\rm eV$ used in the analysis (see Sec.~\ref{sec:analysis})~\footnote{This choice differs from the binning in Fig.~3 of Ref.~\cite{Ruz:2024gkl}, resulting in a different visualization of the experimental data.}.

\subsection{Data analysis}
\label{sec:analysis}

\noindent In this Section, we present the methodology used to derive the bounds obtained in this work. Our approach applies to scenarios in which axions are produced through two distinct channels, namely their coupling to photons, mediated by $g_{a\gamma}$, which also governs the conversions responsible for the final photon signal, and their coupling to an additional particle species $Y$, mediated by $g_{aY}$. The statistical analysis then allows us to derive the combined constraint on the coupling $g_{a\gamma}$ and $g_{aY}$. Throughout this section, we adopt a general formalism, treating $Y$ as an arbitrary particle species. This approach allows for a straightforward specialization to the cases $Y=e$ and $Y=N$, which will be discussed in the following Sections. The differences between these scenarios arise solely from the computation of the specific processes contributing to axion production.

\noindent Axions produced inside the Sun via the Primakoff effect and the coupling with $Y$ induce a photon conversion signal in the hard X-ray range. The expected counts in the $i$-th NuSTAR energy bin, from the $j$-th region ($j=s,\,b$ with $s$ and $b$ denoting the signal and background regions, respectively), and for the $k$-th module $(k=A,\,B)$ are given by

\begin{equation}
    N_{\gamma,i}^{j,k}\left(h,m_a\right)=\Delta t^k\,\int{dE\,A_{\rm eff}^{j,k}\left(E\right)\,\epsilon_{\rm RMF,i}^k\left(E\right)\frac{dN_{a,\rm tot}^j}{dA\,dt\,dE}\left(E\right)
    \,P_{a\rightarrow\gamma}\left(E,h,m_a\right)}\,,
\end{equation}

\noindent where $\Delta t^{k}$ is the exposure time, $A_{\rm eff}^{j,k}$ and $\epsilon_{\rm RMF,i}^k$ are respectively the effective area and the energy response matrix of a given module, and $P_{a\rightarrow\gamma}\left(E,h,m_a\right)\sim g_{a\gamma}^2$ is the axion-photon conversion probability. \noindent The analysis was performed using a bin width of 80 eV and considering the energy range 4
to 15 keV~\footnote{We enlarged the 4-11 keV energy window considered in~\cite{Ruz:2024gkl} to include the axion-conversion signal from  $\prescript{57}{}{\rm Fe}$ (Sec.\ref{sec:axion-nucleon}) de-excitation. This choice has little impact on the constraints in the $g_{a\gamma}$ and $g_{ae}$ parameter space because the axion signal induced by these couplings are strongly peaked at energies $<10$ keV and fall off rapidly at higher energies.
}.

\noindent The differential axion flux $\frac{dN_{a,\rm tot}^j}{dA\,dt\,dE}$ can be decomposed as the sum of a Primakoff and $Y$ contributions

\begin{equation}
    \frac{dN_{a,\rm tot}^j}{dA\,dt\,dE}=\frac{dN_{a,\rm P}^j}{dA\,dt\,dE}+\frac{dN_{a,Y}^j}{dA\,dt\,dE}\,.
\end{equation}

\noindent The Primakoff flux scales with $g_{a\gamma}^2$, whereas the flux originating from $Y$ coupling scales with $g_{aY}^2$. Hence, the photon counts deriving from axion conversion in the Sun can be expressed as

\begin{equation}
\label{eq:gamma from conversion}
N_{\gamma,i}^{j,k}=N_{\gamma P,i}^{j,k}\left(g_{a\gamma}^4\right)+N_{\gamma Y,i}^{j,k} \left(g_{a\gamma}^2\cdot g_{aY}^2\right)\,.
\end{equation}

\noindent The total photon signal detected by a given module, in a given energy bin and region, is the sum of the counts expressed in Eq.~(\ref{eq:gamma from conversion}) and a background contribution $z_{i}^{s,k}$

\begin{equation}
\label{eq:lambda}
    \lambda_{i}^{k}=N_{\gamma P,i}^{j,k}\left(g_{a\gamma}^4\right)+N_{\gamma Y,i}^{j,k} \left(g_{a\gamma}^2\cdot g_{aY}^2\right)+z_{i}^{s,k}\,.
\end{equation}

\noindent Assuming a uniform background of counts within the solar disk, we can infer the background signal from the counts in the background region

\begin{equation}
\label{eq:background}
    z_{i}^{s,k}=\left(t_{i}^k-N_{\gamma P,i}^{b,k}-N_{\gamma Y,i}^{b,k}\right)\frac{A_{\odot}^s}{A_{\odot}^b}\,,
\end{equation}

\noindent where $t_{i}^k$ is the observed signal in the background region, while $A_{\odot}^s$ and $A_{\odot}^b$ are the signal and background areas, respectively.

\noindent Assuming that counts in different energy bins are independent Poisson random variables, the likelihood for a given set of observed counts ${n_i^k}$ in the signal region is given by

\begin{equation}
    \mathcal{L}^k\propto \prod_i\frac{e^{-\lambda_{i}^k}\lambda_{i}^{n_{i}^{k}}}{n_{i}^{k}!}\,.
\end{equation}

\noindent In what follows, the statistical analysis is performed using the combined likelihood, obtained by multiplying the individual likelihoods of the two modules.

\noindent In a region of the parameter space in which the flux of axions from Primakoff effect is subdominant with respect to the flux originating from the interaction with $Y=e,N$, the likelihood is a direct function of the product of the two couplings. Then, we can integrate the Bayesian posterior probability distribution function (PDF) with respect to $\left(g_{a\gamma}^2\cdot g_{aY}^2\right)$, applying the prior $\Theta\left(g_{a\gamma}^2\cdot g_{aY}^2\right)$, with $\Theta\left(x\right)$ denoting the Heaviside theta function, up to the value that includes the 95\% of the PDF area. The result of this procedure is a bound on the product of the couplings, as a function of the axion mass.

More in general, we shall cover also the regime in which neither of the two contributions is dominant. 
For a given axion mass, this can be carried out by fixing a particular value of $g_{aY}$, and integrating the posterior PDF with respect to $g_{a\gamma}^{2}$, using a Bayesian prior $\Theta\left(g_{a\gamma}^2\right)$.
This procedure yields essentially the same bounds as~\cite{Ruz:2024gkl} in the Primakoff-dominated regime, and as the procedure described above when $g_{aY}$ dominates, up to very small differences from the choice of priors.

\section{Results}
\label{sec:results}

\subsection{Bounds on axion-electron coupling}
\label{subsec:axion-electron}

\begin{figure}[H]
    \centering
    \begin{subfigure}[t]{0.49\textwidth}
        \centering
        \includegraphics[width=\textwidth]{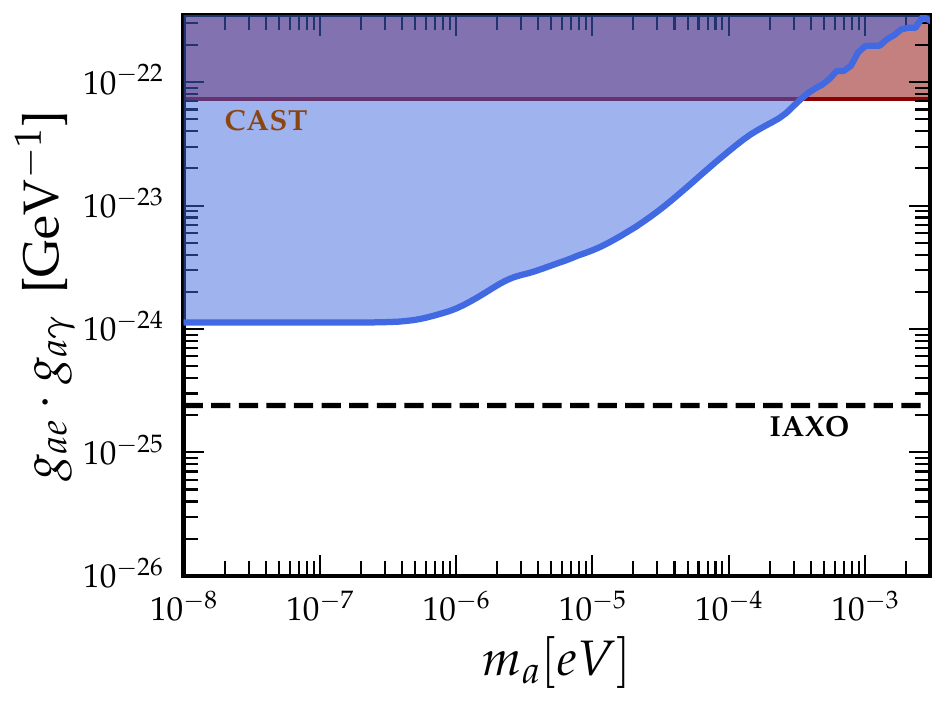}
    \end{subfigure}
    \hfill
    \begin{subfigure}[t]{0.49\textwidth}
        \centering
        \includegraphics[width=\textwidth]{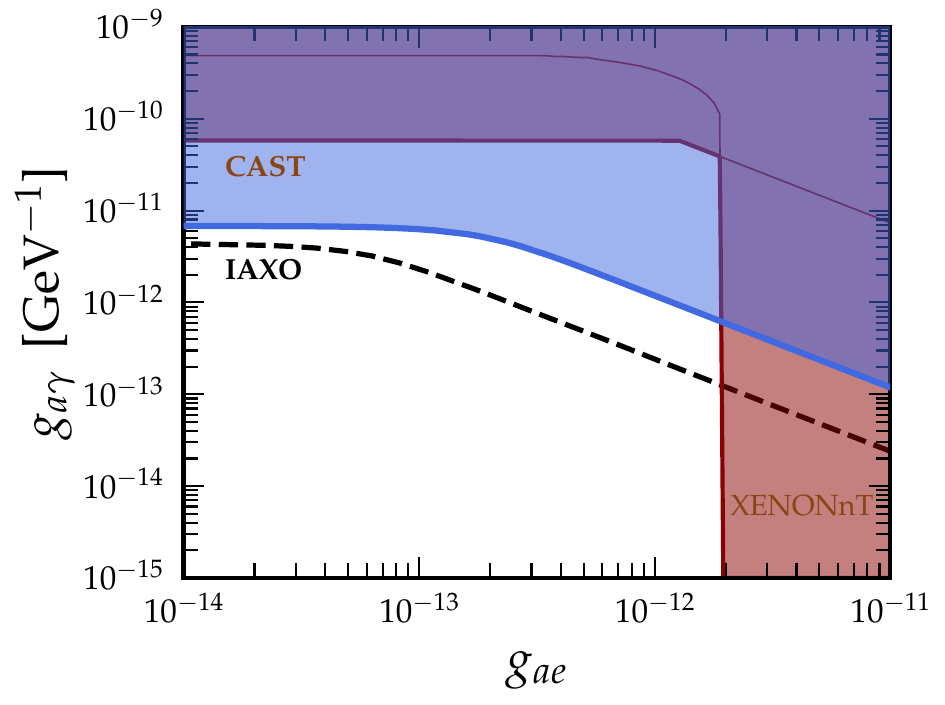}
    \end{subfigure}
    \caption{(\textbf{Left}) 95\% CL bound on the product $g_{ae}\cdot g_{a\gamma}$ obtained in this work (blue region). We also show the limit from the laboratory experiment CAST~\cite{CAST:2025klf}, and the projected sensitivity of IAXO~\cite{IAXO:2019mpb}. (\textbf{Right}) 95\% CL exclusion limit on the $g_{a\gamma}-g_{ae}$ plane (blue region), valid for any $m_a\lesssim10^{-6}\, \rm eV$. The excluded region by CAST is obtained merging the bounds from~\cite{CAST:2024eil,CAST:2025klf}. The excluded region by XENONnT~\cite{XENON:2022ltv}, and the projected sensitivity of IAXO~\cite{IAXO:2019mpb} are also included in the plot.}
    \label{fig:Bounds_experimental_gae}
\end{figure}

The 95\% CL limits on the axion-electron and axion-photon couplings are shown in Fig.~\ref{fig:Bounds_experimental_gae}. As explained in Sec.~\ref{sec:analysis}, when $g_{a\gamma}$ is sufficiently small, Primakoff production in the Sun is subdominant and the solar axion flux is instead controlled by the electron-induced processes proportional to $g_{ae}^2$. In this regime, the analysis of the NuSTAR signal leads to a constraint on the product $g_{ae}\cdot g_{a\gamma}$, shown in the left panel of Fig.~\ref{fig:Bounds_experimental_gae}. It is independent of the axion mass for $m_a\lesssim 10^{-6}\,\mathrm{eV}$, while it weakens at larger masses because of the loss of coherence in the axion-photon conversion probability, as discussed in Ref.~\cite{Ruz:2024gkl}. The left panel of Fig.~\ref{fig:Bounds_experimental_gae} also includes the leading laboratory constraints on $g_{ae}\cdot g_{a\gamma}$, from the CAST experiment~\cite{CAST:2025klf}, together with the projected sensitivity of the future experiment IAXO~\cite{IAXO:2019mpb}. 

\noindent The right panel of Fig.~\ref{fig:Bounds_experimental_gae} shows the 95\% CL bound in the $g_{a\gamma}$–$g_{ae}$ plane, and applies to all axion masses $m_a\lesssim10^{-6}\, \rm eV$. At small $g_{a\gamma}$ this exclusion reproduces the product bound shown in the left panel of Fig.~\ref{fig:Bounds_experimental_gae}, while at larger $g_{a\gamma}$ Primakoff production dominates the solar axion flux and the constraint approaches the NuSTAR Primakoff limit of Ref.~\cite{Ruz:2024gkl}. For comparison, we also show the XENONnT constraint from solar-axion searches with electronic recoils, which currently provides the strongest laboratory
limit on $g_{ae}$ alone~\cite{XENON:2022ltv}.

As illustrated in Fig.~\ref{fig:Bounds_experimental_gae}, the NuSTAR limits probe a substantial region of the axion parameter space that is not accessible to current laboratory experiments. Additional competitive, but more model-dependent, astrophysical constraints are discussed in Appendix~\ref{app:Astro Bounds}.

\subsection{Bounds on axion-nucleon coupling}
\label{subsec:axion-electron}

\begin{figure}[H]
    \centering
    \begin{subfigure}[t]{0.49\textwidth}
        \centering
        \includegraphics[width=\textwidth]{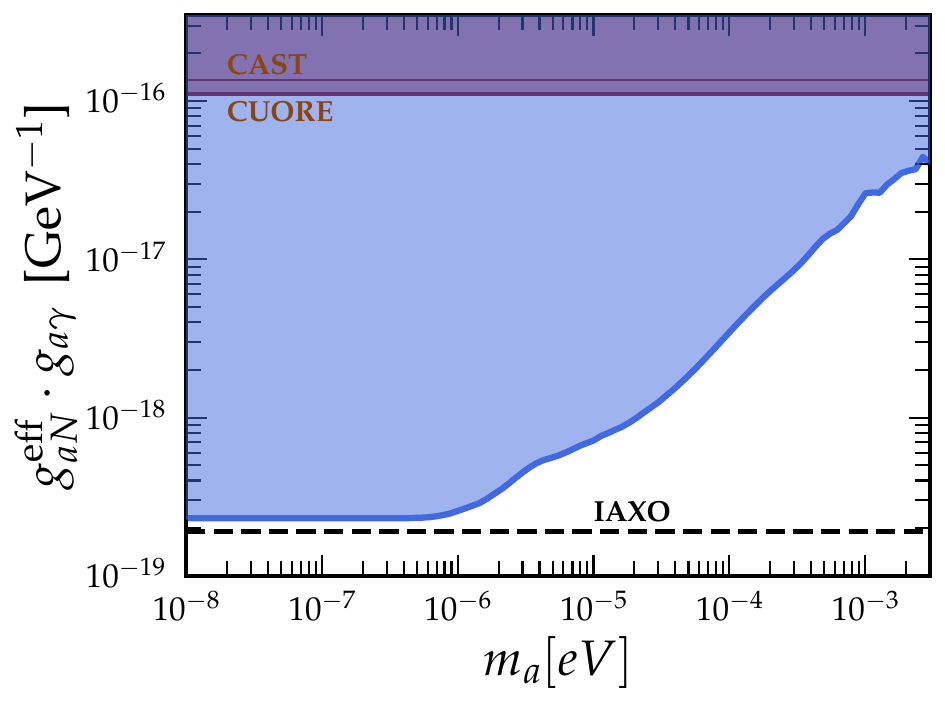}
    \end{subfigure}
    \hfill
    \begin{subfigure}[t]{0.49\textwidth}
        \centering
        \includegraphics[width=\textwidth]{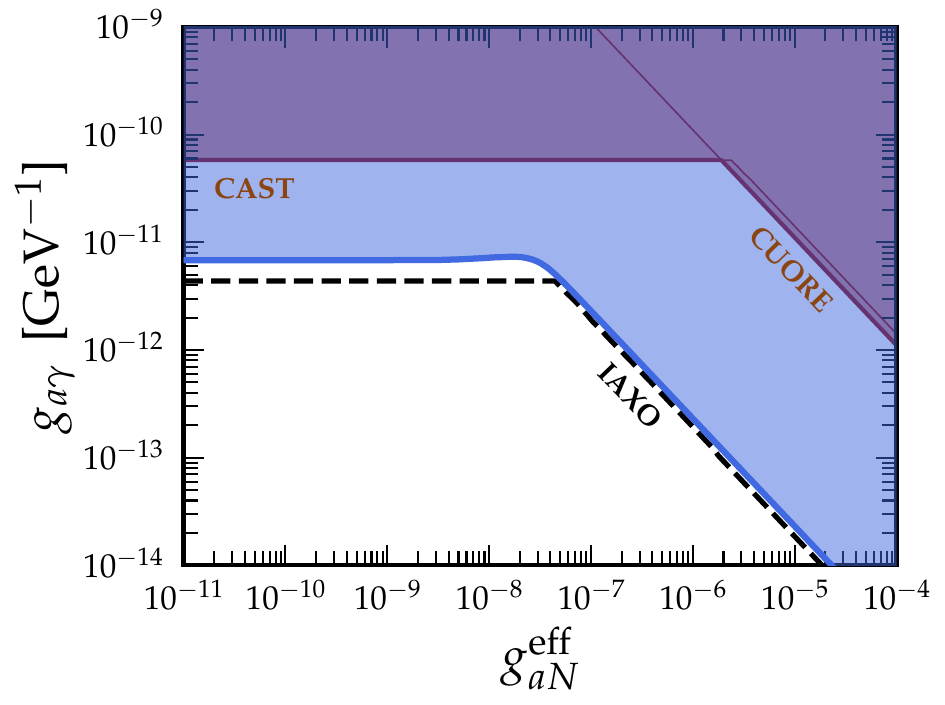}
    \end{subfigure}
    \caption{(\textbf{Left}) 95\% CL bound on the product $g_{aN}^{\rm eff}\cdot g_{a\gamma}$ obtained in this work (blue region).
   We also show
   the limits from the laboratory experiments
   CAST~\cite{CAST:2009jdc} and CUORE~\cite{Li:2015tyq}, and the projected sensitivity of IAXO~\cite{DiLuzio:2021qct}. (\textbf{Right}) 95\% CL exclusion limit on the $g_{a\gamma}-g_{aN}^{\rm eff}$ plane (blue region), valid for any $m_a\lesssim10^{-6}\, \rm eV$. The CAST bound is a combination of the bounds from~\cite{CAST:2024eil} and~\cite{CAST:2009jdc}. The other lines show the bound from CUORE~\cite{Li:2015tyq}, and the projected sensitivity of IAXO~\cite{DiLuzio:2021qct}.}
    \label{fig:Bounds_experimental_gaN}
\end{figure}

The NuSTAR limits on the product $g_{aN}^{\rm eff}\cdot g_{a\gamma}$ are shown in the left panel of Fig.~\ref{fig:Bounds_experimental_gaN}, applicable when Primakoff production is subdominant. The bound in the $g_{a\gamma}-g_{aN}^{\rm eff} $ plane, which applies to axion masses $m_a\lesssim10^{-6}\, \rm eV$, is shown in the right panel of the same figure. We compare these results to the most stringent laboratory constraints, namely those from CAST~\cite{CAST:2024eil,CAST:2009jdc} and the underground experiment CUORE, which searched for solar axions via the coherent inverse Primakoff process~\cite{Li:2015tyq}.

At the qualitative level, the main features of the NuSTAR limits are similar to those already discussed for the axion-electron coupling. We emphasize that the bounds derived in this work significantly improve over existing laboratory constraints, approaching the projected sensitivity of the future IAXO for $m_a\lesssim10^{-6}\, \rm eV$.
Astrophysical constraints are presented in Appendix~\ref{app:Astro Bounds}.

\section{Conclusions}
\label{sec:Conclusion}

Solar axion searches have long been among the most sensitive probes of light, weakly interacting pseudoscalars. While laboratory helioscopes such as CAST rely on artificial magnetic fields to convert solar axions into detectable X-rays, the magnetic field of the solar atmosphere itself provides a natural and far more extended conversion region, turning X-ray satellites pointed at the quiet Sun into powerful axion observatories. Building on this idea, we have extended the analysis of Ref.~\cite{Ruz:2024gkl} beyond the purely Primakoff scenario to include the additional production channels mediated by the axion-electron and axion-nucleon couplings.

\noindent Using the long quiet-Sun NuSTAR exposure obtained on 21 February 2020, during a quiet-Sun period near solar minimum, we have derived 95\% CL upper limits on the products of the relevant couplings, $g_{ae}\cdot g_{a\gamma}\lesssim 1.1\times 10^{-24}\,\rm GeV^{-1}$ and $g_{aN}^{\rm eff}\cdot g_{a\gamma}\lesssim 2.3\times 10^{-19}\,\rm GeV^{-1}$, valid for $m_a\lesssim 10^{-6}\,\rm eV$, together with bounds in the full $(g_{a\gamma},\,g_{ae})$ and $(g_{a\gamma},\,g_{aN}^{\rm eff})$ planes that smoothly interpolate between the Primakoff-dominated regime and the BCA- and nuclear-dominated ones. The inclusion of the $14.4\,\rm keV$ line from M1 transitions of $\prescript{57}{}{\rm Fe}$ adds a nearly monochromatic feature, well separated from the smoother Primakoff and electron-induced continua, that would constitute a clean spectroscopic signature in the event of a future positive detection.

\noindent These constraints probe a substantial region of parameter space that is currently inaccessible to ground-based experiments, improving over the leading limits on the relevant coupling products from CAST~\cite{CAST:2025klf,CAST:2024eil,CAST:2009jdc} and CUORE~\cite{Li:2015tyq} in the mass range $m_a\lesssim 10^{-6}\,\rm eV$, and approaching the projected reach of IAXO~\cite{IAXO:2019mpb,DiLuzio:2021qct} in the nucleon channel.

\noindent A number of astrophysical observations formally reach into deeper parts of the same parameter space (Appendix~\ref{app:Astro Bounds}), but they hinge on the detailed modelling of stellar structure, evolution, magnetic-field morphology, and plasma density in systems far less well constrained than our own Sun. Our bounds therefore offer a robust, systematics-controlled complement to those probes.

\noindent The 2020 NuSTAR campaign will likely remain the most suitable dataset of this kind until the next solar minimum, expected around 2030, when an extended quiet-Sun programme with NuSTAR or successor hard-X-ray missions could substantially increase exposure and energy reach. Together with IAXO and the next generation of underground and X-ray observatories, high-precision observations of the quiet Sun are likely to play a central role in the search for axions and axion-like particles.

\section*{Acknowledgements}

We gratefully acknowledge the support of the NuSTAR operations, software, and calibration teams in the execution and analysis of these observations. 

MR, MT and TZ acknowledge support from the Research grant TAsP (Theoretical Astroparticle Physics) funded by \textsc{infn}. The work of MT is supported by the Italian Ministry of University and Research (MUR) via the PRIN 2022 Project No. 2022F2843 CUP I53D23000670006 “Addressing systematic uncertainties in searches for dark matter” funded by MIUR.
The work of MR and TZ is supported by the European Union – Next Generation EU and by the Italian Ministry of University and Research (MUR) via the PRIN 2022 project n. 20228WHTYC - CUP D53C24003550006. TZ acknowledges support from the UniTO-IBS Avogadro Programme (collaboration agreement between the University of Torino and the Institute for Basic Science, South Korea).
ET is supported by the European Union’s Horizon 2020
research and innovation programme under the Marie Skłodowska-Curie grant
agreement No 101204903 (APARAX).

MG acknowledges support from the Spanish Agencia Estatal de Investigación under grant PID2019-108122GB-C31, funded by MCIN/AEI/10.13039/501100011033, and from the “European Union NextGenerationEU/PRTR” (Planes complementarios, Programa de Astrofísica y Física de Altas Energías). He also acknowledges support from grant PGC2022-126078NB-C21, “Aún más allá de los modelos estándar,” funded by MCIN/AEI/10.13039/501100011033 and “ERDF A way of making Europe.” Additionally, MG acknowledges funding from the European Union’s Horizon 2020 research and innovation programme under the European Research Council (ERC) grant agreement ERC-2017-AdG788781 (IAXO+). 

JKV, FRC, and JR acknowledge support from the Deutsche Forschungsgemeinschaft (DFG, German Research Foundation) under Germany’s Excellence Strategy – Cluster of Excellence “Color meets Flavor”, EXC 3107 – Project-ID 533766364. JKV and FRC also acknowledge funding from the German federal and state program "Professorinnenprogramm 2030" Project-ID 01FP24167Q.

\appendix
\section{Comparison to astrophysical bounds}
\label{app:Astro Bounds}

\begin{figure}[H]
    \centering
    \begin{subfigure}[t]{0.49\textwidth}
        \centering
        \includegraphics[width=\textwidth]{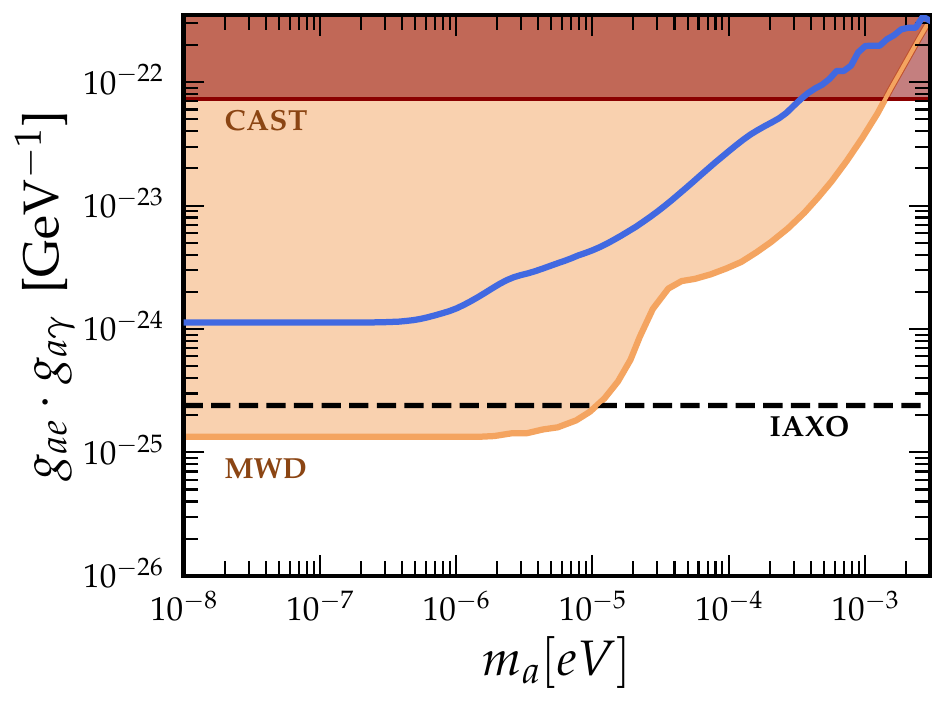}
    \end{subfigure}
    \hfill
    \begin{subfigure}[t]{0.49\textwidth}
        \centering
        \includegraphics[width=\textwidth]{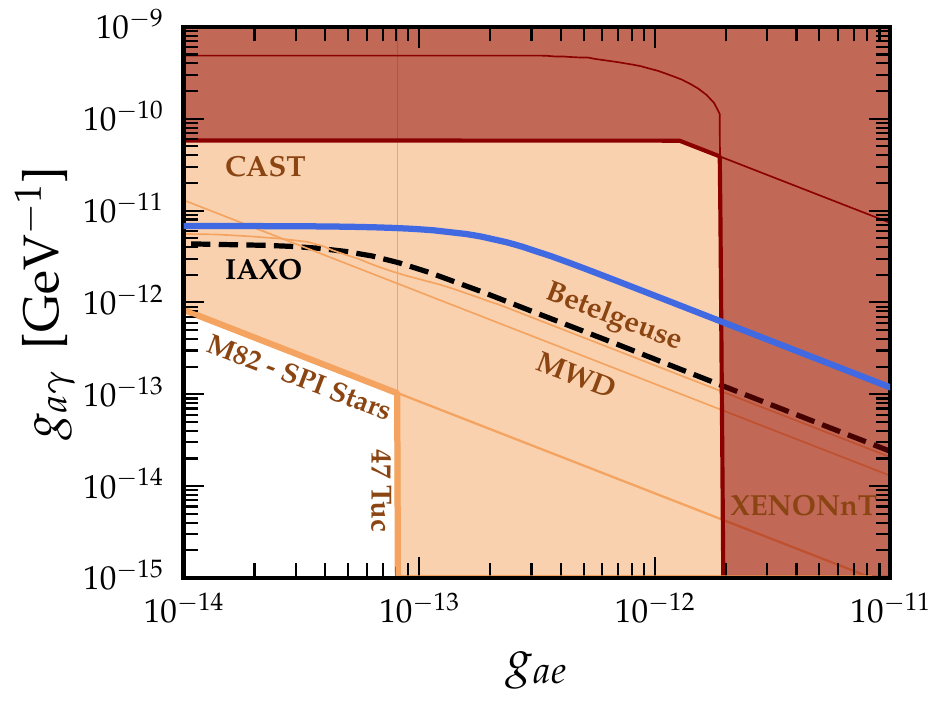}
    \end{subfigure}
    \caption{(\textbf{Left}) Comparison between the 95\% CL bound on the product $g_{ae}\cdot g_{a\gamma}$ obtained in this work (blue line),
   the limits and sensitivities from laboratory experiments shown in the left panel of Fig.~\ref{fig:Bounds_experimental_gae},
   and the astrophysical bound obtained from axion-photon conversions in magnetic white dwarfs (MWD)~\cite{Dessert:2021bkv}. (\textbf{Right}) Comparison of our bound in the $g_{a\gamma} - g_{ae}$ plane (blue line) with 
   the limits and sensitivities from laboratory experiments shown in the right panel of Fig.~\ref{fig:Bounds_experimental_gae} and with
    astrophysical constraints. 
    In this plot we assume $m_a\lesssim10^{-6}\rm eV.$
    The astrophysical constraints are based on observations of Betelgeuse~\cite{Xiao:2022rxk}, M82~\cite{Ning:2025tit}, nearby pre-supernova stars (SPI Stars)~\cite{Mittal:2025xwt},
    magnetic white dwarfs (MWD)~\cite{Dessert:2021bkv},
    and from white dwarf cooling in 47 Tucanae~\cite{Fleury:2025ahw}.
    The limits from~\cite{Ning:2025tit,Xiao:2022rxk,Mittal:2025xwt} apply to axion masses $m_a\lesssim10^{-11}\, \rm eV$.
    The limits from M82 and from pre-supernova stars (we show the strongest bounds presented in~\cite{Mittal:2025xwt}) are very similar and therefore appear as a single line.}
    \label{fig:Bounds astrophysical gae}
\end{figure}

\begin{figure}[H]
    \centering
    \begin{subfigure}[t]{0.49\textwidth}
        \centering
        \includegraphics[width=\textwidth]{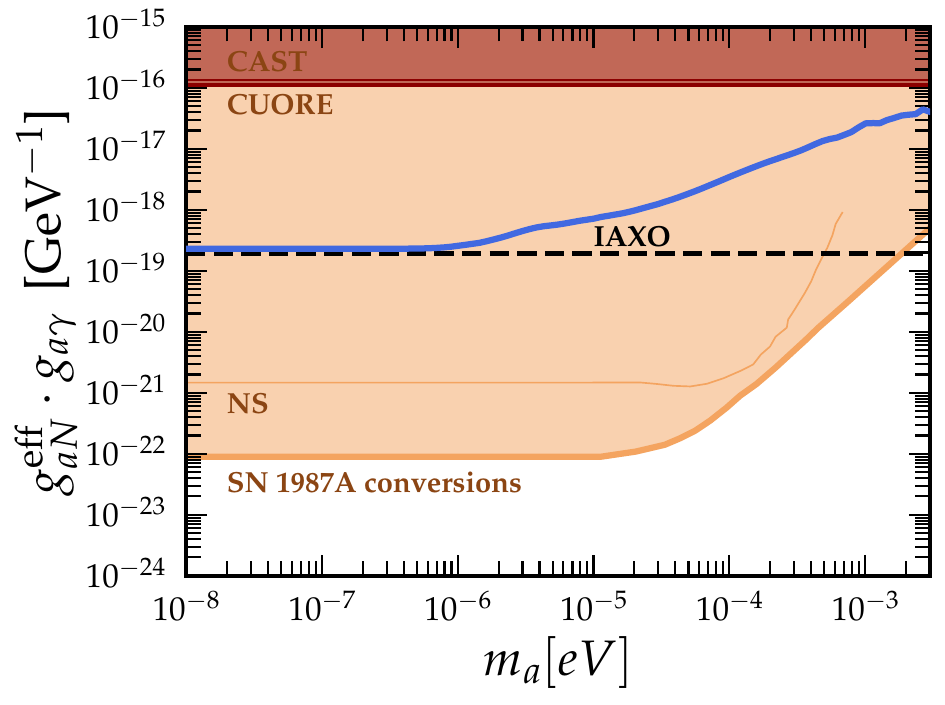}
    \end{subfigure}
    \hfill
    \begin{subfigure}[t]{0.49\textwidth}
        \centering
        \includegraphics[width=\textwidth]{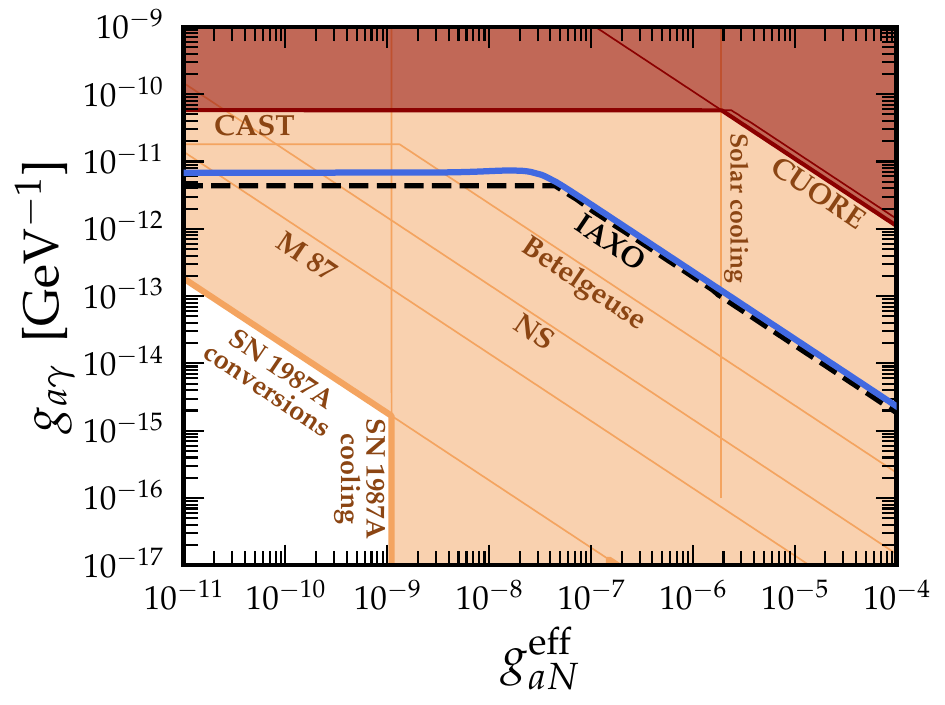}
    \end{subfigure}
    \caption{(\textbf{Left}) Comparison of the 95\% CL bound on the product $g_{aN}^{\rm eff}\cdot g_{a\gamma}$ obtained in this work (blue line) with the limits and sensitivities from laboratory experiments shown in the left panel of Fig.~\ref{fig:Bounds_experimental_gaN} and with astrophysical bounds. The astrophysical limits shown are from neutron stars (NS)~\cite{Ning:2025fqd} and from SN1987A, derived from the results of~\cite{Manzari:2024jns} (\textbf{Right}) Comparison of the bound in the $g_{a\gamma} - g_{aN}^{\rm eff}$ plane obtained in this work (blue line), with the limits and sensitivities from laboratory experiments shown in the right panel of Fig.~\ref{fig:Bounds_experimental_gaN} and with other astrophysical probes. In this plot we assume $m_a\lesssim10^{-6}\,\rm eV.$ The astrophysical bounds are based on searches of axion-photon conversion signals from SN1987A~\cite{Manzari:2024jns}, neutron stars (NS)~\cite{Ning:2025fqd}, Betelgeuse~\cite{Candon:2025vpv}, M87~\cite{Ning:2025kyu}, and from the cooling of the Sun and SN1987A~\cite{DiLuzio:2021qct}. The limits from~\cite{Candon:2025vpv,Ning:2025kyu} apply to axion masses $m_a\lesssim10^{-11}-10^{-10}\, \rm eV$. The constraints from~\cite{Manzari:2024jns} shown here apply to $m_a\lesssim10^{-10}\, \rm eV$, while for larger masses they reduce to the ones in the left panel of Fig.\ref{fig:Bounds astrophsysical gaN}.}
    \label{fig:Bounds astrophsysical gaN}
\end{figure}

In this Appendix we present strong, although model-dependent, astrophysical constraints that apply in the parameter space of Figs.~\ref{fig:Bounds astrophysical gae},~\ref{fig:Bounds astrophsysical gaN}.

In the left panel of Fig.~\ref{fig:Bounds astrophysical gae}, we consider astrophysical probes that are sensitive to the combination of couplings $g_{ae}\cdot g_{a\gamma}$. The strongest limits arise from X-ray observations of magnetic white-dwarf stars~\cite{Dessert:2021bkv}. In these systems, axions can be produced in the core of the stars through electron bremsstrahlung and then converted to X-rays in the magnetosphere.
In the right panel of Fig.~\ref{fig:Bounds astrophysical gae}, in addition to these limits, we also show the constraints derived from NuSTAR X-ray observations of the Betelgeuse red supergiant star~\cite{Xiao:2022rxk}, the galaxy M82~\cite{Ning:2025tit} and from INTEGRAL/SPI gamma-ray observations of a sample of nearby pre-supernova stars~\cite{Mittal:2025xwt}.
The bounds in~\cite{Xiao:2022rxk,Mittal:2025xwt} have been obtained considering the production of axions via $g_{ae}$ and $g_{a\gamma}$ couplings in the stars and their conversion to photons in the magnetic field of our Galaxy (in the right panel of Fig.~\ref{fig:Bounds astrophysical gae} we report the tightest limits presented in these references). 
Ref.~\cite{Ning:2025tit} computed the axion flux produced by stars of M82 (via $g_{ae}$ and $g_{a\gamma}$) and the conversion of axions into photons in the magnetic fields of M82.
The limits from Beltelgeuse, M82 and SPI observations of stars shown in the right panel of Fig.~\ref{fig:Bounds astrophysical gae} apply to axion masses $m_a\lesssim10^{-11}\, \rm eV$ 
and they  weaken significantly for larger values, therefore they are not shown in the left panel, as they are subdominant to the limits from magnetic white dwarf stars in this region of parameter space.
Axions can also be constrained through observations of stellar properties, since their production in stars provides an additional exotic cooling channel which can affect the stellar evolution.
In the right panel of Fig.~\ref{fig:Bounds astrophysical gae} we show the limits obtained from the cooling of white dwarfs in the globular cluster 47 Tucanae~\cite{Fleury:2025ahw}. Similar but slightly weaker limits have been reported from the observations of white dwarfs in the Gaia DR3 sample~\cite{Alberino:2026yxi}, and from red-giant branch stars~\cite{Capozzi:2020cbu,Straniero:2020iyi}.

Moving to the axion-nucleon coupling, in the left panel of Fig.~\ref{fig:Bounds astrophsysical gaN} we show the constraints on the product $g_{aN}^{\rm eff}\cdot g_{a\gamma}$ from observations of the Supernova SN1987A, based on axion-photon conversion in the magnetic fields of the progenitor star~\cite{Manzari:2024jns}. 
See also~\cite{Grifols:1996id,Brockway:1996yr,Payez:2014xsa,Hoof:2022xbe,Calore:2020tjw,Candon:2025fnb} for bounds on photon fluxes from axion conversion derived from SN1987A and the diffuse supernova flux.
We also report limits from searches of axion-photon conversion signals from neutron star populations~\cite{Ning:2025fqd}.
In the right panel of Fig.~\ref{fig:Bounds astrophsysical gaN}, in addition to these limits, we show bounds from observations of Betelgeuse~\cite{Candon:2025vpv} and the galaxy M87~\cite{Ning:2025kyu}. They apply to axion masses $m_a\lesssim10^{-11}-10^{-10}\, \rm eV$ and they weaken for larger values, therefore they are not shown in the right panel.
Limits on $g_{aN}^{\rm eff}$ can be derived from the cooling of stars. In the right panel of Fig.~\ref{fig:Bounds astrophsysical gaN} we show the constraints derived in~\cite{DiLuzio:2021qct} from solar observations and from the cooling of SN1987A (using the analysis in~\cite{Carenza:2019pxu}).

In conclusion, several competitive astrophysical bounds complement those derived in this work. However, as emphasized in~\cite{Ruz:2024gkl} and as can be deduced from the previous discussion, their systematic uncertainties are typically larger than those of the NuSTAR limits presented here. 
In fact, modeling axion production and conversion in these astrophysical systems is often hindered by significant uncertainties in their structure and evolution, as well as in environmental properties such as magnetic field strengths and plasma densities.
In contrast, the analysis presented here exploits the Sun as a target, a system whose properties are comparatively well known, allowing for a more robust evaluation of model uncertainties.

\newpage

\bibliographystyle{bibi}
\bibliography{bibliography.bib}

\end{document}